\documentclass[a4paper,fleqn]{cas-dc}

\usepackage[numbers]{natbib}
\newcommand{\Rmnum}[1]{\uppercase\expandafter{\romannumeral #1}}  %定义命令输入大写罗马数字
\usepackage{balance} % 加载 balance 宏包
\usepackage{titlesec}
\titleformat*{\section}{\small \bf}
\titleformat*{\subsection}{\small \em}
\titleformat*{\subsubsection}{\small \em}
\usepackage{CJKutf8} % 转义字符
\usepackage{amsmath,amssymb,amsfonts}
\usepackage{amsthm}
\usepackage{algorithmic}

\usepackage{tabularx}
\usepackage{booktabs}

\usepackage{bbding}
\usepackage[caption=false,font=normalsize,labelfont=sf,textfont=sf]{subfig}
\usepackage{multirow}
\usepackage{makecell}
\usepackage{textcomp}
\usepackage{stfloats}
\usepackage{url}
\usepackage{verbatim}
\usepackage{graphicx}
\usepackage{amssymb} % 三角等号
\usepackage[ruled, vlined, linesnumbered]{algorithm2e}
\usepackage{tabularx}
\usepackage{booktabs}
\usepackage[dvipsnames, svgnames, x11names]{xcolor}

\usepackage{array}
\usepackage{bbding}
\usepackage[caption=false,font=normalsize,labelfont=sf,textfont=sf]{subfig}
\usepackage[T1]{fontenc} %摘要字体加粗
\usepackage{amssymb} % 三角等号
\usepackage{threeparttable}

\begin{document}
\let\WriteBookmarks\relax
\def\floatpagepagefraction{1}
\def\textpagefraction{.001}
\shorttitle{Semantic Anchor Communication System for Point Cloud Transmission in Internet of Vehicles}

\title [mode = title]{Semantic Anchor Communication System for Point Cloud Transmission in Internet of Vehicles}

\author[1]{Ruixing Ren}[style=chinese,
orcid=0009-0000-1183-9512]
\ead{renruixing0604@163.com}

\affiliation[1]{organization={School of Electronic and Information Engineering},
	addressline={Beijing Jiaotong University}, 
	city={Beijing},
	postcode={100044}, 
	country={China}}

\affiliation[2]{organization={School of Communications and Information Engineering},
	addressline={Nanjing University of Posts and Telecommunications},
	city={Nanjing},
	postcode={210003}, 
	country={China}}

\author[1]{Junhui Zhao}[style=chinese,
orcid=0000-0001-5958-6622]
\cormark[1]
\ead{junhuizhao@hotmail.com}

\author[1]{Wei Zhao}[style=chinese, orcid=0009-0004-9202-150X]
\ead{25120164@bjtu.edu.cn}

\author[1]{Gaoxin Zhang}[style=chinese, orcid=0009-0007-9670-4975]
\ead{g72815767@hotmail.com}

\author[2]{Xiaoming Wang}[style=chinese]
\ead{xmwang@njupt.edu.cn}

\cortext[cor1]{Corresponding author}

\begin{abstract}
Joint transmission of point cloud geometry and semantics under bandwidth constraints is required for cooperative perception in the Internet of Vehicles (IoV). Per-point semantics cannot be explicitly preserved by existing geometric compression methods, while interpretable three-dimensional point sets cannot be directly reconstructed via task-feature communication. The paper proposes a spatial semantic anchor communication system. Distinct from global codewords and task features, anchors with explicit coordinates are employed as transmission units. Local shape descriptors and category labels are bound to a unified spatial reference, preserving the correspondence between local position, shape and semantics. At the transmitter, anchor representations are extracted and then converted into digital frames through fixed-point quantization and lossless compression. At the receiver, labeled point clouds are reconstructed through anchor-constrained local expansion, and pose-aligned multi-view observations are fused. On the 500-frame SemanticPOSS test set, the results demonstrate that spatial semantic anchors enable interpretable joint reconstruction and fusion of geometry and semantics with low transmission overhead.
\end{abstract}

\begin{keywords}
Internet of Vehicles (IoV) \sep point cloud \sep semantic communication \sep semantic anchor \sep collaborative perception
\end{keywords}

\maketitle

% =========================================================================
\section{Introduction}

Environmental perception by single-vehicle LiDAR is limited by occlusion and field-of-view constraints \cite{RenITS,ZhaoISCC}. Cooperative perception in the Internet of Vehicles (IoV) compensates for perceptual blind spots via shared multi-view observations \cite{How2comm,MRCNet,CoSDH,RenUAV}, yet substantial wireless resources are consumed by continuous transmission of voluminous raw point cloud data. To support three-dimensional (3D) environment reconstruction at the receiver, both geometric structure and semantic categories of point clouds must be preserved in transmitted representations. Accordingly, the problem of labeled point cloud reconstruction and multi-view fusion under limited transmission budgets is investigated in this paper.

Rate-distortion performance of geometry or attributes is primarily optimized by existing learning-based point cloud compression methods \cite{DeepPCC,RENO}, whereas intermediate features for detection or segmentation are mostly transmitted in cooperative perception schemes \cite{CORE,CoSDH}.
Per-point semantics are generally not explicitly preserved by the former, while labeled 3D point sets for spatial fusion are not directly output by the latter.
Semantic communication opens up new avenues for task-relevant information transmission \cite{LYU,RenGANImage}. However, the position-category correspondence of local structures is barely retained by scene-level descriptions, and data prior dependency is readily induced in generative decoders.

To this end, a semantic anchor communication system for point cloud transmission in the IoV is proposed. At the transmitter, local position, shape and category are jointly represented via spatial anchors, and quantized codewords are encapsulated into digital frames. At the receiver, labeled point clouds are reconstructed under anchor constraints, and successfully received multi-view observations are fused in a unified coordinate system. The core contributions of the paper are summarized below.
\begin{itemize}
	\item A spatial semantic anchor coding method is proposed to explicitly preserve the position-shape-category correspondence in compact codewords, enabling joint geometry-semantic transmission.
	\item An anchor-constrained local expansion and multi-view fusion method is devised, where point cloud reconstruction is guided by transmitted coordinates and complementary observations are integrated via pose alignment, semantic matching and temporal weighting.
	\item Experiments are conducted on the SemanticPOSS dataset with zero-bit decoding and equal-byte raw coordinate transmission as baselines, to verify the effectiveness of spatial semantic anchors and analyze the roles of semantic supervision and multi-view fusion.
\end{itemize}

The remainder of this paper is organized as follows. Related work is reviewed in Section \ref{SecRW}. The system model is presented in Section \ref{Sec2}. Algorithm design is detailed in Section \ref{Sec3}. Simulation results are reported in Section \ref{Sec4}. Conclusions are drawn in Section \ref{Sec5}.

\section{Related Work}\label{SecRW}
\subsection{Point Cloud Compression and Generative Reconstruction}

Point cloud compression is typically centered on rate-distortion performance of geometry or attributes. Code streams are structured by conventional G-PCC/V-PCC via octrees, surface approximation or 2D projection, while spatiotemporal correlation and entropy models are further exploited in learning-based methods. Only latent residuals were encoded via inter-frame prediction as proposed by Akhtar et al. \cite{Akhtar2024}. Geometry and attributes were modeled by DeepPCC via joint multi-scale neighborhoods, sparse convolutions and local self-attention \cite{DeepPCC}. Onboard inference overhead was reduced by RENO through multi-scale sparse tensors and occupancy prediction \cite{RENO}, and a joint voxel Transformer and sparse convolution framework was devised by TSC-PCAC for point cloud attributes \cite{TSC}. Faithful recovery of geometry or attributes is primarily pursued by such methods, while the binding between spatial position, local shape and semantic categories is generally not explicitly maintained.

Incomplete observations are complemented by generative reconstruction via data priors. 3D completion was formulated as a conditional diffusion process in DiffComplete \cite{DiffComplete}. Dynamic neural fields and multi-node perception-based reconstruction synthesize continuous environments by leveraging spatiotemporal priors or node collaboration. \cite{Zheng,LuB}. Missing structures can be recovered by such methods, yet outputs may rely more on data priors than current frame information at low bitrates. This paper constrains local generation with spatial anchors and employs a zero-bit reference to verify whether the reconstruction is genuinely driven by the current frame's codeword.

\subsection{Semantic Communication and Cooperative Perception in IoV}
Delivery of task-relevant information rather than bit-wise reproduction is prioritized in semantic communication. Architectures, evaluation metrics and application scenarios of semantic V2X have been investigated in existing studies \cite{LYU}, and vehicular semantic communication has been optimized from perspectives of knowledge bases, multi-user services and multimodal tasks \cite{Xia,FengSemCom}. Task accuracy and network utility are commonly emphasized in related evaluations, whereas practical serialization of point cloud representations, frame-level link outages and per-point semantic reconstruction are rarely addressed.

Information selection and fusion across vehicles are further explored in cooperative perception. Benchmarks for multi-vehicle or vehicle-infrastructure cooperative perception in real traffic were provided by datasets such as V2V4Real, V2X-Real and TUMTraf-V2X \cite{V2V4Real,V2XReal,TUMTraf}. For communication efficiency, bird's-eye view representations were recovered by CORE via intermediate feature compression and reconstruction supervision \cite{CORE}. Cooperation overhead was reduced by How2comm, MRCNet and bandwidth modeling methods via latency-aware interaction, motion-robust communication and link budget constraints respectively \cite{How2comm,MRCNet,BandwidthEff}. Cross-model collaboration for open heterogeneous agents was investigated by HEAL, Hetecooper and GenComm \cite{HEAL,Hetecooper,GenComm}. Cooperation regions were selected and intermediate features and detection results were fused by CoSDH based on supply-demand relations \cite{CoSDH}. Bird's-eye view features or detection results for downstream detection are primarily transmitted by such methods, while labeled 3D point sets for direct geometric error evaluation are generally not output at the receiver.

Overall, distinct objectives are pursued by geometric compression, task-feature communication and generative completion respectively, yet no unified representation enables serializable transmission, labeled point cloud reconstruction and spatial fusion. Spatial semantic anchors are adopted in this paper to bridge transmitter encoding, digital links and receiver reconstruction. Codeword validity, geometry-semantic tradeoffs and transmission overhead are examined respectively via zero-bit baseline, equal-byte raw coordinate transmission and anchor count experiments.

\begin{figure}[t]
	\centerline{\includegraphics[width=3.5in,keepaspectratio]{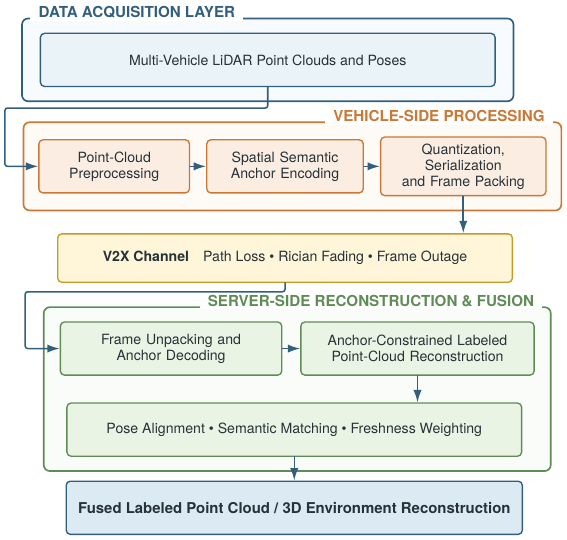}}
	\caption{System Model}
	\label{Fig1}
\end{figure}

\section{System Model}\label{Sec2}
As illustrated in Fig.~\ref{Fig1}, a three-layer hierarchical architecture encompassing data acquisition, data processing and transmission, and environment reconstruction is employed. The set of vehicles is denoted as \(\mathcal{V}=\{v_1,v_2,\dots,v_N\}\). The raw point cloud acquired by vehicle \(v_i\) at time $t$ is expressed as
\begin{equation}
	\mathcal{P}^{(t)}_i=\left\{ \mathbf{p}^{(t)}_{i,1},\mathbf{p}^{(t)}_{i,2},\dots,\mathbf{p}^{(t)}_{i,M_i} \right\}
	\label{eq:P}
\end{equation}
where \(M_i=|\mathcal{P}_i^{(t)}|\) denotes the number of points, and each point \(\mathbf{p}^{(t)}_{i,j}\in\mathbb{R}^d\) carries multi-dimensional features including spatial coordinates and reflection intensity. For notational simplicity, the time superscript $t$ is omitted hereafter.

\subsection{Data Acquisition Layer}
Surrounding point clouds are acquired in real time by each vehicle via onboard LiDAR, and data are exchanged over V2V/V2I links including DSRC and C-V2X. The union of point clouds acquired by all vehicles at the same instant is expressed as
\begin{equation}
	\mathcal{P}=\bigcup_{i=1}^{N}\mathcal{P}_i
	\label{eq:union}
\end{equation}

\subsection{Data Processing and Transmission Layer}
Preprocessed point cloud \(\tilde{\mathcal{P}}_i\) is derived from raw point cloud \(\mathcal{P}_i\) via denoising, downsampling and normalization
\begin{equation}
	\tilde{\mathcal{P}}_i=\Phi_{\text{pre}}(\mathcal{P}_i)
	\label{eq:pre}
\end{equation}
where \(\Phi_{\text{pre}}(\cdot)\) denotes the preprocessing operator. Then, hierarchical features are extracted via PointNet++ and retained as a set of spatial anchor codewords rather than aggregated into a global vector
\begin{equation}
	\mathbf{T}_i=\Phi_{\text{enc}}(\tilde{\mathcal{P}}_i)
	=\left\{ \left( \mathbf{a}_{i,k},\ \mathbf{t}_{i,k},\ y_{i,k} \right) \right\}_{k=1}^{A}
	\label{eq:enc}
\end{equation}
where \(\mathbf{a}_{i,k}\in\mathbb{R}^3\) is the normalized coordinate of the $k$-th anchor, \(\mathbf{t}_{i,k}\in\mathbb{R}^{d_t}\) is its local shape codeword, \(y_{i,k}\) is the anchor semantic category, and \(A\ll|\tilde{\mathcal{P}}_i|\). The anchor set is processed via quantization, serialization, lossless compression and frame encapsulation to form digital frames
\begin{equation}
	\mathbf{c}_i=\Phi_{\text{sc}}(\mathbf{T}_i)
	\label{eq:sc}
\end{equation}
where \(\Phi_{\text{sc}}(\cdot)\) denotes the described semantic encoding and frame encapsulation process, and \(\mathbf{c}_i\) is the digital frame for transmission, which is delivered to the server over the V2X channel.

\subsection{V2X Channel Model}
IoV communications are characterized by high mobility, rapid time variation and multipath propagation \cite{RenRIS}. A V2X channel model comprising large-scale path loss and small-scale Rician fading is adopted.

The complex channel gain of the link between vehicle \(v_i\) and the receiver is expressed as
\begin{equation}
	h_i=\frac{\tilde{h}_i}{\sqrt{L(d_i)}},
	\label{eq:gain}
\end{equation}
where \(d_i\) is the transceiver distance, \(L(d_i)\) is the linear-scale path loss, and \(\tilde{h}_i\) is the small-scale fading coefficient.

Path loss follows the 3GPP TR 37.885 model \cite{Song}
\begin{equation}
	PL_{\mathrm{dB}}(d_i)
	=32.4+20\log_{10}(f_c)+30\log_{10}(d_i),
	\label{eq:PL}
\end{equation}
where \(f_c\) is the carrier frequency in GHz and \(d_i\) is measured in meters. The corresponding linear path loss is given by
\begin{equation}
	L(d_i)=10^{PL_{\mathrm{dB}}(d_i)/10}.
	\label{eq:PLlinear}
\end{equation}

Small-scale fading follows the Rician model
\begin{equation}
	\tilde{h}_i=
	\sqrt{\frac{K}{K+1}}\,h_{\mathrm{LOS}}
	+\sqrt{\frac{1}{K+1}}\,h_{\mathrm{NLOS}},
	\label{eq:rician}
\end{equation}
where $K$ is the Rician factor representing the power ratio of line-of-sight to scattered paths, and \(h_{\mathrm{LOS}}\) and \(h_{\mathrm{NLOS}}\) denote the line-of-sight and non-line-of-sight components respectively.

Given transmit power \(P_t\), channel bandwidth $B$ and noise power spectral density \(N_0\), the instantaneous received signal-to-noise ratio (SNR) of vehicle \(v_i\) is given by
\begin{equation}
	\gamma_i=\frac{|h_i|^2P_t}{BN_0}.
	\label{eq:snr}
\end{equation}

The corresponding theoretical achievable transmission rate is given by
\begin{equation}
	R_i=B\log_2(1+\gamma_i).
	\label{eq:rate}
\end{equation}

To characterize communication outages caused by deep fading, we adopt a frame-availability model based on the instantaneous received SNR. When the instantaneous received SNR $\gamma_i$ of the link associated with vehicle $v_i$ falls below a threshold $\gamma_{\mathrm{th}}$, the corresponding frame is considered unavailable; otherwise, its codewords are assumed to arrive intact at the receiver. This model captures the impact of link outages on point cloud reconstruction and multi-source fusion without explicitly simulating the physical-layer encoding and decoding processes.

\subsection{Environment Reconstruction Layer}
Successfully received vehicle codewords are decoded at the server. The reconstructed point cloud \(\hat{\mathcal{P}}_i\) and corresponding per-point semantic labels \(\hat{\mathbf{y}}_i\) are given by
\begin{equation}
	\hat{\mathcal{P}}_i,\ \hat{\mathbf{y}}_i=\Phi_{\text{dec}}(\mathbf{c}_i)
	\label{eq:dec}
\end{equation}
where \(\Phi_{\text{dec}}(\cdot)\) consists of an anchor expansion decoder and a per-point semantic prediction head. The set of vehicles with successfully received codewords is denoted as \(\mathcal{V}_{\mathrm{rx}}\subseteq\mathcal{V}\), with \(K=|\mathcal{V}_{\mathrm{rx}}|\). Further, multi-source point cloud fusion is expressed as
\begin{equation}
	\mathcal{P}_{\mathrm{fused}}
	=\Psi_{\mathrm{fuse}}
	\left(
	\left\{\hat{\mathcal{P}}_i\mid i\in\mathcal{V}_{\mathrm{rx}}\right\}
	\right).
\end{equation}
where \(\Psi_{\text{fuse}}(\cdot)\) is the fusion operator, which jointly accounts for spatial alignment, feature matching and weight allocation.

\section{Algorithm Design}\label{Sec3}
Preprocessing, spatial anchor coding and frame encapsulation, receiver-side anchor expansion and per-point semantic prediction, and multi-vehicle point cloud fusion methods are detailed sequentially.

\subsection{Data Preprocessing}\label{subsec:pre}
Raw point cloud \(\mathcal{P}_i\) acquired by vehicle \(v_i\) is input to data preprocessing, and normalized point cloud \(\tilde{\mathcal{P}}_i\) is produced. The entire processing pipeline is expressed as
\begin{equation}
	\mathcal{P}_i
	\xrightarrow{\Phi_{\mathrm{crop}}}
	\mathcal{P}_i^{\mathrm{crop}}
	\xrightarrow{\Phi_{\mathrm{sor}}}
	\mathcal{P}_i^{\mathrm{clean}}
	\xrightarrow{\Phi_{\mathrm{vox}}}
	\mathcal{P}_i^{\mathrm{down}}
	\xrightarrow{\Phi_{\mathrm{norm}}}
	\tilde{\mathcal{P}}_i,
	\label{eq:preflow}
\end{equation}
where \(\Phi_{\mathrm{crop}}\), \(\Phi_{\mathrm{sor}}\), \(\Phi_{\mathrm{vox}}\) and \(\Phi_{\mathrm{norm}}\) denote distance cropping, statistical outlier filtering, voxel downsampling and coordinate normalization respectively.

\textbf{(1) Distance cropping and denoising.} Points within the effective sensing range of the sensor are retained via distance cropping to yield \(\mathcal{P}_i^{\mathrm{crop}}\). The mean distance \(\bar{d}_{i,j}\) between the $j$-th point and its \(K_{\mathrm{nbr}}\) nearest neighbors is then computed, where \(K_{\mathrm{nbr}}\) is the neighborhood size. The mean and standard deviation of average neighbor distances across all points in the frame are denoted as \(\mu_{d,i}\) and \(\sigma_{d,i}\) respectively. A point is classified as an outlier and removed when
\begin{equation}
	\bar{d}_{i,j}>
	\mu_{d,i}+\lambda_{\mathrm{sor}}\sigma_{d,i}
	\label{eq:sor}
\end{equation}
holds, where \(\lambda_{\mathrm{sor}}\) is the outlier detection threshold coefficient. The denoised point cloud is denoted as \(\mathcal{P}_i^{\mathrm{clean}}\).

\textbf{(2) Voxel downsampling.} 3D space is partitioned into voxels of side length \(r_{\mathrm{voxel}}\) by the operator \(\Phi_{\mathrm{vox}}\). One representative point is retained per occupied voxel to produce the downsampled point cloud \(\mathcal{P}_i^{\mathrm{down}}\).

\textbf{(3) Coordinate normalization.} The centroid \(\boldsymbol{\mu}_i\) and normalization scale \(s_i\) of \(\mathcal{P}_i^{\mathrm{down}}\) are computed as
\begin{equation}
	\boldsymbol{\mu}_i=
	\frac{1}{|\mathcal{P}_i^{\mathrm{down}}|}
	\sum_{\mathbf{p}\in\mathcal{P}_i^{\mathrm{down}}}\mathbf{p},
	\qquad
	s_i=
	\max_{\mathbf{p}\in\mathcal{P}_i^{\mathrm{down}}}
	\left\|\mathbf{p}-\boldsymbol{\mu}_i\right\|.
	\label{eq:norm}
\end{equation}
The normalized point cloud is expressed as
\begin{equation}
	\tilde{\mathcal{P}}_i
	=
	\left\{
	\frac{\mathbf{p}-\boldsymbol{\mu}_i}{s_i}
	\ \middle|\
	\mathbf{p}\in\mathcal{P}_i^{\mathrm{down}}
	\right\}.
	\label{eq:normpoint}
\end{equation}
The centroid \(\boldsymbol{\mu}_i\) and scale \(s_i\) are transmitted with codewords to enable metric coordinate recovery of reconstructed point clouds at the receiver. The normalized point cloud is sampled to a fixed size for encoder input, with oversampling applied when point count is insufficient. The sampled point cloud remains denoted as \(\tilde{\mathcal{P}}_i\), with \(M_{\mathrm{in}}=\left|\tilde{\mathcal{P}}_i\right|\).

\subsection{Hierarchical Feature Extraction}\label{subsec:encoder}
The preprocessed point cloud \(\tilde{\mathcal{P}}_i\) is fed into hierarchical feature extraction, and the spatial semantic anchor codeword set \(\mathbf{T}_i\) is generated. An $L$-layer PointNet++ set abstraction architecture \cite{Qi} is adopted for the encoder. Its processing pipeline is expressed as
\begin{equation}
	\left(
	\mathcal{X}_i^{(0)},
	\mathcal{F}_i^{(0)}
	\right)
	\xrightarrow{\mathrm{SA}_1}
	\cdots
	\xrightarrow{\mathrm{SA}_L}
	\left(
	\mathcal{X}_i^{(L)},
	\mathcal{F}_i^{(L)}
	\right)
	\xrightarrow{\Phi_{\mathrm{anchor}}}
	\mathbf{T}_i,
	\label{eq:encflow}
\end{equation}
where \(\mathcal{X}_i^{(0)}=\tilde{\mathcal{P}}_i\) is the encoder input coordinate set, \(\mathcal{F}_i^{(0)}\) is the corresponding input point feature set, $L$ is the number of set abstraction layers, and \(\Phi_{\mathrm{anchor}}\) is the anchor codeword generation module.

For the \(\ell\)-th set abstraction layer, its input and output satisfy
\begin{equation}
	\left(
	\mathcal{X}_i^{(\ell)},
	\mathcal{F}_i^{(\ell)}
	\right)
	=
	\mathrm{SA}_{\ell}
	\left(
	\mathcal{X}_i^{(\ell-1)},
	\mathcal{F}_i^{(\ell-1)}
	\right),
	\quad \ell=1,\ldots,L,
	\label{eq:sa}
\end{equation}
where \(\mathcal{X}_i^{(\ell)}\) and \(\mathcal{F}_i^{(\ell)}\) are the output coordinate set and feature set of the \(\ell\)-th layer respectively. \(M_\ell\) representative points are obtained via farthest point sampling in this module. For each representative point, \(K_\ell\) neighboring points are selected within radius \(r_\ell\). Then, local features are then aggregated via shared multilayer perceptrons, max pooling and channel attention.

The final layer omits global pooling and retains $A$ anchors along with their local features
\begin{equation}
	\mathcal{X}_i^{(L)}
	=
	\left\{\mathbf{a}_{i,k}\right\}_{k=1}^{A},
	\qquad
	\mathcal{F}_i^{(L)}
	=
	\left\{\mathbf{f}_{i,k}\right\}_{k=1}^{A},
	\label{eq:anchorfeatures}
\end{equation}
where $A$ is the anchor count, and \(\mathbf{a}_{i,k}\) and \(\mathbf{f}_{i,k}\) are the coordinate and local feature of the $k$-th anchor respectively. The anchor codeword generation module maps \(\mathbf{f}_{i,k}\) to local shape code \(\mathbf{t}_{i,k}\) and predicts anchor category \(y_{i,k}\), yielding
\begin{equation}
	\mathbf{T}_i
	=
	\left\{
	\left(
	\mathbf{a}_{i,k},
	\mathbf{t}_{i,k},
	y_{i,k}
	\right)
	\right\}_{k=1}^{A}.
	\label{eq:anchorcode}
\end{equation}
Preservation of anchor-level representations avoids local spatial position information loss induced by global pooling.

\subsection{Semantic Encoding and Frame Encapsulation}\label{subsec:frame}
The spatial semantic anchor set \(\mathbf{T}_i\) serves as input to semantic encoding and frame encapsulation, from which transmittable digital frames \(\mathbf{c}_i\) are produced. Its processing pipeline is expressed as
\begin{equation}
	\mathbf{T}_i
	\xrightarrow{\Phi_{\mathrm{quant}}}
	\mathbf{Q}_i
	\xrightarrow{\Phi_{\mathrm{pack}}}
	\mathbf{c}_i,
	\label{eq:frameflow}
\end{equation}
where \(\Phi_{\mathrm{quant}}\) and \(\Phi_{\mathrm{pack}}\) denote anchor quantization and frame encapsulation operations respectively, and \(\mathbf{Q}_i\) is the quantized anchor set.

For the $k$-th anchor, signed fixed-point quantization is applied to its coordinates and local shape codes, and fixed-length integer encoding is adopted for semantic categories
\begin{equation}
	\mathbf{q}_{i,k}
	=
	\left[
	Q_b(\mathbf{a}_{i,k}),
	Q_b(\mathbf{t}_{i,k}),
	E_b(y_{i,k})
	\right],
	\label{eq:anchorquant}
\end{equation}
where \(Q_b(\cdot)\) is a b-bit uniform quantization operator, \(E_b(\cdot)\) is the semantic category encoding operator, and \(\mathbf{q}_{i,k}\) is the $k$-th quantized anchor. All quantized anchors form
\begin{equation}
	\mathbf{Q}_i=\left\{\mathbf{q}_{i,k}\right\}_{k=1}^{A}.
\end{equation}

Subsequently, \(\mathbf{Q}_i\) is serialized, losslessly compressed and concatenated with a frame header
\begin{equation}
	\mathbf{c}_i
	=
	\operatorname{Pack}
	\left(
	\mathcal{H}_i,\,
	\operatorname{Zlib}
	\left(
	\operatorname{Serialize}(\mathbf{Q}_i)
	\right)
	\right),
	\label{eq:pack}
\end{equation}
where \(\mathcal{H}_i\) is the frame header recording protocol version, codeword dimensions, quantization parameters, point cloud centroid \(\boldsymbol{\mu}_i\) and normalization scale \(s_i\).

\subsection{Anchor-Based Point Cloud Reconstruction}\label{subsec:dec}
Successfully received digital frames \(\mathbf{c}_i\) are fed into the receiver, and the reconstructed point cloud \(\hat{\mathcal{P}}_i\) and its per-point semantic labels \(\hat{\mathbf y}_i\) are produced. The decoding pipeline is expressed as
\begin{equation}
	\mathbf{c}_i
	\xrightarrow{\Phi_{\mathrm{unpack}}}
	\hat{\mathbf T}_i
	\xrightarrow{\Phi_{\mathrm{dec}}}
	\left(
	\hat{\mathcal{P}}_i,
	\hat{\mathbf y}_i
	\right),
	\label{eq:decflow}
\end{equation}
where \(\Phi_{\mathrm{unpack}}\) denotes unpacking, lossless decompression and dequantization operations, and \(\hat{\mathbf T}_i\) is the recovered anchor set.

For the $k$-th anchor, its coordinate $\hat{\mathbf a}_{i,k}$, local shape code $\hat{\mathbf t}_{i,k}$ and category embedding are concatenated and mapped to an anchor feature
\begin{equation}
	\mathbf h_{i,k}
	=
	\Phi_{\mathrm{lift}}
	\left(
	\left[
	\hat{\mathbf a}_{i,k},
	\hat{\mathbf t}_{i,k},
	\operatorname{Emb}(\hat y_{i,k})
	\right]
	\right),
	\label{eq:lift}
\end{equation}
where $\operatorname{Emb}(\cdot)$ is the category embedding function and $\Phi_{\mathrm{lift}}$ is the feature lifting module. After interaction through residual blocks and self-attention modules, local sub-points are generated from anchor features by an offset prediction module.

Given $U$ sub-points expanded per anchor, the $j$-th sub-point generated by the $k$-th anchor is given by
\begin{equation}
	\hat{\mathbf p}_{i,k,j}
	=
	\tanh
	\left(
	\hat{\mathbf a}_{i,k}
	+
	\rho\,\mathbf o_{i,k,j}
	\right),
	\qquad
	j=0,\ldots,U-1,
	\label{eq:expand}
\end{equation}
where $\mathbf o_{i,k,j}$ is the bounded local offset output by the offset prediction module and $\rho$ is the offset scale. Let $\mathbf o_{i,k,0}=\mathbf 0$ to ensure at least one zero-offset sub-point per group. As the output passes through $\tanh$, this sub-point lies at $\tanh(\hat{\mathbf a}_{i,k})$ and does not strictly coincide with the quantized anchor coordinate. The reconstructed point cloud is given by
\begin{equation}
	\hat{\mathcal P}_i
	=
	\left\{
	\hat{\mathbf p}_{i,k,j}
	\mid
	k=1,\ldots,A,\;
	j=0,\ldots,U-1
	\right\}.
	\label{eq:reconstructedset}
\end{equation}
With $U$ sub-points generated per anchor, the point count of the reconstructed point cloud is
\begin{equation}
	M_{\mathrm{rec}}
	=
	\left|\hat{\mathcal P}_i\right|
	=
	AU.
	\label{eq:recpoints}
\end{equation}
Each sub-point inherits the semantic category of its parent anchor
\begin{equation}
	\hat y_{i,k,j}=\hat y_{i,k}.
	\label{eq:labelinherit}
\end{equation}
Accordingly, the decoder outputs semantic labels corresponding one-to-one to reconstructed points while recovering the geometric structure of the point cloud.

\subsection{Multi-source Point Cloud Fusion}\label{subsec:fuse}
Successfully decoded vehicle point clouds, per-point semantic labels and auxiliary information are fed into multi-source point cloud fusion. Fused point clouds and semantic labels in a unified coordinate system are produced as
\begin{equation}
	\left(
	\mathcal P_{\mathrm{fused}},
	\mathbf y_{\mathrm{fused}}
	\right)
	=
	\Psi_{\mathrm{fuse}}
	\left(
	\left\{
	\hat{\mathcal P}_i,
	\hat{\mathbf y}_i,
	\mathbf G_i,
	q_i,
	t_i^{\mathrm{obs}}
	\right\}_{i\in\mathcal V_{\mathrm{rx}}}
	\right),
	\label{eq:fusionflow}
\end{equation}
where \(\mathcal V_{\mathrm{rx}}\) is the set of vehicles with successfully received codewords, \(\mathbf G_i\) is the pose of vehicle \(v_i\), \(q_i\) is the frame reception indicator, and \(t_i^{\mathrm{obs}}\) is the observation timestamp.

\textbf{(1) Spatial alignment.} Let \(\mathbf G_i=(\mathbf R_i,\mathbf u_i)\), where \(\mathbf R_i\) and \(\mathbf u_i\) are the rotation matrix and translation vector from the vehicle coordinate system to the unified coordinate system respectively. Reconstructed points are first restored to metric coordinates using the centroid \(\boldsymbol{\mu}_i\) and scale \(s_i\) from the frame header, and then transformed to the unified coordinate system
\begin{equation}
	\mathbf p_{i,j}^{\mathrm{world}}
	=
	\mathbf R_i
	\left(
	s_i\hat{\mathbf p}_{i,j}
	+
	\boldsymbol{\mu}_i
	\right)
	+
	\mathbf u_i.
	\label{eq:align}
\end{equation}

\textbf{(2) Overlapping point matching.} Per-point semantic categories obtained from decoding are represented as one-hot descriptors \(\phi(\cdot)\). For points \(\mathbf p\) and \(\mathbf q\) from different vehicles, they are classified as overlapping observations of the same region when
\begin{equation}
	\|\mathbf p-\mathbf q\|<r_m
	\quad\land\quad
	\cos
	\left(
	\phi(\mathbf p),
	\phi(\mathbf q)
	\right)
	\geq\tau_m
	\label{eq:match}
\end{equation}
holds, where \(r_m\) is the geometric matching radius and \(\tau_m\) is the descriptor similarity threshold. With one-hot descriptors, this similarity criterion is equivalent to a category consistency check.

\textbf{(3) Weighted fusion.} The fusion weight of vehicle \(v_i\) is determined jointly by link availability and observation timeliness
\begin{equation}
	w_i
	=
	q_i
	\exp
	\left(
	-\frac{|\Delta t_i|}{\tau_f}
	\right),
	\label{eq:weight}
\end{equation}
where \(q_i\) is the frame reception indicator, taking value 1 for successful reception and 0 for erasure. \(\Delta t_i\) is the time difference between the observation and the reference instant, and \(\tau_f\) is the temporal decay coefficient. For matched observations, positions are averaged with normalized weights, and semantic categories are determined via weighted voting. Unmatched points are selectively retained by weight to complement scene coverage.

\subsection{Loss Function}\label{subsec:loss}
The preprocessed point cloud \(\tilde{\mathcal P}_i\), reconstructed point cloud \(\hat{\mathcal P}_i\), anchor ground-truth categories and encoder-output class probabilities serve as inputs to the loss function, from which a scalar loss for model training is derived. The training objective is composed of geometric reconstruction loss and anchor semantic classification loss
\begin{equation}
	L_{\mathrm{total}}
	=
	\beta L_{\mathrm{rec}}
	+
	\gamma L_{\mathrm{sem}},
	\label{eq:total}
\end{equation}
where \(\beta\) and \(\gamma\) are the weights of geometric reconstruction loss and semantic classification loss respectively.

\textbf{(1) Geometric reconstruction loss.} As point clouds are unordered sets, a bidirectional Chamfer-L1 distance is adopted to quantify the geometric discrepancy between \(\hat{\mathcal P}_i\) and \(\tilde{\mathcal P}_i\)
\begin{equation}
	\begin{aligned}
		L_{\mathrm{rec}}
		={}&
		\frac{1}{|\hat{\mathcal P}_i|}
		\sum_{\hat{\mathbf p}\in\hat{\mathcal P}_i}
		\min_{\mathbf p\in\tilde{\mathcal P}_i}
		\left\|
		\hat{\mathbf p}-\mathbf p
		\right\|_2
		\\
		&+
		\frac{1}{|\tilde{\mathcal P}_i|}
		\sum_{\mathbf p\in\tilde{\mathcal P}_i}
		\min_{\hat{\mathbf p}\in\hat{\mathcal P}_i}
		\left\|
		\mathbf p-\hat{\mathbf p}
		\right\|_2 .
	\end{aligned}
	\label{eq:rec}
\end{equation}
This loss employs unsquared Euclidean nearest-neighbor distances, hence the designation Chamfer-L1. No pre-established point index correspondence is required for point clouds.

\textbf{(2) Anchor semantic classification loss.} Sampling indices are retained by set abstraction modules during farthest point sampling. By compounding sampling indices of all set abstraction layers sequentially, final anchors can be precisely mapped to corresponding points in the preprocessed point cloud, and anchor ground-truth categories are retrieved.

Let the ground-truth category of the $k$-th anchor be \(y_{i,k}^{*}\), and the class probability distribution output by the encoder classification head be \(\boldsymbol{\pi}_{i,k}\). The valid anchor set is defined as
\begin{equation}
	\mathcal K_i
	=
	\left\{
	k\mid y_{i,k}^{*}\neq y_{\mathrm{ignore}}
	\right\},
	\label{eq:validanchors}
\end{equation}
where \(y_{\mathrm{ignore}}\) denotes the category identifier excluded from supervision. The semantic classification loss is given by
\begin{equation}
	L_{\mathrm{sem}}
	=
	-\frac{
		\sum_{k\in\mathcal K_i}
		\omega_{y_{i,k}^{*}}
		\log \pi_{i,k,y_{i,k}^{*}}
	}{
		\sum_{k\in\mathcal K_i}
		\omega_{y_{i,k}^{*}}
	},
	\label{eq:sem}
\end{equation}
where \(\pi_{i,k,y_{i,k}^{*}}\) is the predicted probability of the k-th anchor belonging to its ground-truth category, and \(\omega_{y_{i,k}^{*}}\) is the loss weight for the corresponding category.

\section{Simulation Experiments}\label{Sec4}
\subsection{Experimental Setup}

\textbf{(1) Dataset and preprocessing.} The SemanticPOSS dataset \cite{SemanticPOSS} is adopted. Three sequences are used for training, and one sequence each is selected for validation and testing. The distance cropping range, statistical filtering neighborhood size, threshold coefficient and voxel size are set to 50 m, 20, 2.0 and 0.05 m respectively. Each frame is sampled to 9216 points. Each point contains seven-dimensional features including normalized coordinates, reflection intensity and coordinates mapped to \([0,1]\).

\textbf{(2) Encoder and decoder configuration.} A three-level PointNet++ set abstraction architecture is adopted for the encoder, with parameters listed in Table \ref{tab:network}. All neighborhood radii are defined in the normalized coordinate space.

\begin{table}[t]
	\centering
	\caption{Encoder network parameters}
	\label{tab:network}
	\footnotesize
	\begin{tabular}{c c c c l}
		\toprule
		Layer &  \(M_\ell\) & \(r_\ell\) &
		\(K_\ell\) & MLP channels \\
		\midrule
		SA1 & 4096 & 0.1 & 16 & $[64,64,128]$ \\
		SA2 & 1024 & 0.2 & 16 & $[128,128,256]$ \\
		SA3 & 512  & 0.4 & 16 & $[256,512,1024]$ \\
		\bottomrule
	\end{tabular}
\end{table}

In this work, \(A=512\), \(d_t=2\) and \(U=18\) are specified, resulting in a reconstruction point count \(M_{\mathrm{rec}}=9216\). The decoder backbone feature dimension is 256, and the initial value of the learnable offset scale \(\rho\) is 0.03.

\textbf{(3) Quantization and frame encapsulation configuration.} 
Anchor coordinates, local shape codes and category identifiers are represented as 16-bit signed integers. The frame header records protocol version, codeword dimensions, quantization parameters, centroid and normalization scale, with the detailed structure listed in Table \ref{tab:frameconfig}. Quantized payloads are losslessly compressed via Zlib. Measured frame lengths and Compression Ratios (CR) are presented in Section \ref{subsec:main}.

\begin{table}[t]
	\centering
	\caption{Anchor codeword and frame encapsulation configuration}
	\label{tab:frameconfig}
	\setlength{\tabcolsep}{1pt}
	\footnotesize
	\begin{tabular}{l c c}
		\toprule
		Field & Elements per anchor & Bytes per anchor \\
		\midrule
		Normalized coordinates & 3 & 6~B \\
		Local shape codes & 2 & 4~B \\
		Semantic category identifier & 1 & 2~B \\
		\midrule
		Anchor count & \multicolumn{2}{c}{$A=512$} \\
		Raw payload after quantization & \multicolumn{2}{c}{6144~B} \\
		Frame header length & \multicolumn{2}{c}{51~B} \\
		\bottomrule
	\end{tabular}
\end{table}

\textbf{(4) Training strategy.} The model is trained with the Adam optimizer, an initial learning rate of \(10^{-4}\), and a mini-batch size of 4. An effective batch size of 32 is achieved via 8-step gradient accumulation, with a total of 100 training epochs. Loss weights are set to \(\beta=0.7\) and \(\gamma=0.2\). The geometric loss is computed by sampling 2048 points each from the target and reconstructed point clouds, while full point clouds are used for evaluation. Semantic category weights are set as the inverse square root of training set frequencies and normalized, with zero weight assigned to unlabeled classes. All checkpoints are selected based on validation mIoU, except for the ablation model with \(\gamma=0\), which is selected based on validation Chamfer Distance (CD).

\begin{figure}[t]
	\centering
	\includegraphics[width=\columnwidth,keepaspectratio]{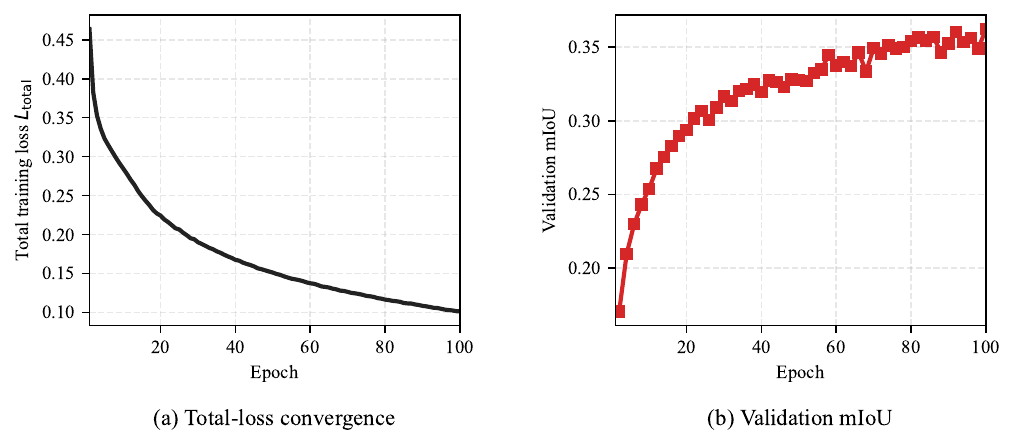}
	\caption{Training convergence of the final model: (a) Total training loss; (b) Validation mIoU.}
	\label{FigTrain}
\end{figure}

Figure \ref{FigTrain} illustrates the training process of the final model. The total loss decreases continuously with a notably slower rate in later stages, and the validation mIoU stabilizes in late training.

\textbf{(5) V2X channel setup.} Default parameters of the V2X channel are listed in Table \ref{tab:params}. Default experiments adopt a received signal-to-noise ratio of 20 dB and zero additional packet loss rate.

\begin{table}[t]
	\centering
	\caption{Default V2X channel parameters}
	\label{tab:params}
	\footnotesize
	\begin{tabular}{l l}
		\toprule
		Parameter & Default value \\
		\midrule
		Carrier frequency \(f_c\) & 5.9 GHz \\
		Transceiver distance d & 50 m \\
		Rician factor K & 7.0 \\
		Channel bandwidth B & 10 MHz \\
		Transmit power \(P_t\) & 0.2 W \\
		Noise power spectral density \(N_0\) & \(10^{-20}\) W/Hz \\
		Received signal-to-noise ratio & 20 dB \\
		Additional packet loss rate & 0 \\
		Frame availability threshold \(\gamma_{\mathrm{th}}\) & 10 dB \\
		\bottomrule
	\end{tabular}
\end{table}

\textbf{(6) Multi-view fusion setup.} To evaluate the complementary coverage provided by reconstructed point clouds, we construct multi-view observations using point clouds captured by the same vehicle at different times in SemanticPOSS. Vehicle motion results in different observation positions and spatial coverage across frames, while maintaining partial overlap. Fusion parameters are set to \(r_m=0.5\) m, \(\tau_m=0.3\) and \(\tau_f=0.2\) s. The candidate point count before fusion is \(K M_{\mathrm{rec}}\), where \(K=|\mathcal V_{\mathrm{rx}}|\). The fused result is resampled to \(M_{\mathrm{rec}}\) points, and spatial alignment uses poses provided by the dataset.

\textbf{(7) Experimental platform.} All experiments are conducted on a single NVIDIA RTX 4090 (24 GB). The software environment consists of Python 3.12.3, PyTorch 2.8.0 and CUDA 12.8.

\subsection{Evaluation Metrics}\label{subsec:metric}
Let the reference and reconstructed point clouds of the $i$-th frame be denoted respectively as
\begin{equation}
	S_i^{\mathrm{ref}}=\tilde{\mathcal P}_i,
	\qquad
	S_i^{\mathrm{rec}}=\hat{\mathcal P}_i,
\end{equation}

\textbf{(1) CD.} Bidirectional mean squared nearest-neighbor distance is adopted to quantify geometric error
\begin{equation}
	\begin{aligned}
		\mathrm{CD}
		\left(
		S_i^{\mathrm{ref}},
		S_i^{\mathrm{rec}}
		\right)
		={}&
		\frac{1}{|S_i^{\mathrm{ref}}|}
		\sum_{\mathbf p\in S_i^{\mathrm{ref}}}
		\min_{\hat{\mathbf p}\in S_i^{\mathrm{rec}}}
		\left\|
		\mathbf p-\hat{\mathbf p}
		\right\|_2^2
		\\
		&+
		\frac{1}{|S_i^{\mathrm{rec}}|}
		\sum_{\hat{\mathbf p}\in S_i^{\mathrm{rec}}}
		\min_{\mathbf p\in S_i^{\mathrm{ref}}}
		\left\|
		\hat{\mathbf p}-\mathbf p
		\right\|_2^2 .
	\end{aligned}
	\label{eq:cd}
\end{equation}
A lower CD indicates better performance. Chamfer-L1 from Equation \eqref{eq:rec} is used for training, while Chamfer-L2 from Equation \eqref{eq:cd} is used for evaluation. Metric CD values are computed after coordinate recovery.

\textbf{(2) F-score.} Given a matching threshold \(\delta\), precision and recall are defined as
\begin{equation}
	\begin{aligned}
		\mathrm{Precision}(\delta)
		&=
		\frac{
			\left|
			\left\{
			\hat{\mathbf p}\in S_i^{\mathrm{rec}}
			\mid
			\min_{\mathbf p\in S_i^{\mathrm{ref}}}
			\|\hat{\mathbf p}-\mathbf p\|_2<\delta
			\right\}
			\right|
		}{
			|S_i^{\mathrm{rec}}|
		},
		\\
		\mathrm{Recall}(\delta)
		&=
		\frac{
			\left|
			\left\{
			\mathbf p\in S_i^{\mathrm{ref}}
			\mid
			\min_{\hat{\mathbf p}\in S_i^{\mathrm{rec}}}
			\|\mathbf p-\hat{\mathbf p}\|_2<\delta
			\right\}
			\right|
		}{
			|S_i^{\mathrm{ref}}|
		}.
	\end{aligned}
	\label{eq:pr}
\end{equation}
The corresponding F-score is given by
\begin{equation}
	\mathrm{F}(\delta)
	=
	\frac{
		2\,\mathrm{Precision}(\delta)\,\mathrm{Recall}(\delta)
	}{
		\mathrm{Precision}(\delta)+\mathrm{Recall}(\delta)
	}.
	\label{eq:f1}
\end{equation}
Main experiments adopt \(\delta=0.01\), \(0.02\) and \(0.05\), corresponding to mean metric values of 0.505, 1.010 and 2.526 m respectively. Fusion experiments directly use metric thresholds. A higher F-score indicates better performance.

\begin{table*}[t]
	\centering
	\caption{Test results under default configuration.}
	\label{tab:main}
	\footnotesize
	\resizebox{0.9\textwidth}{!}{%
		\begin{tabular}{l c c c c}
			\toprule
			Method or baseline
			& Frame size (B)
			& CD
			& F@0.01
			& mIoU \\
			\midrule
			Zero-bit baseline
			& 0
			& 0.119053
			& 0.0035
			& 0.0033 \\
			Equal-byte coordinate direct transmission
			& 5353
			& 0.000250
			& 0.4531
			& -- \\
			\textbf{Proposed method}
			& 5358
			& 0.001715$\pm$0.000057
			& 0.2906$\pm$0.0004
			& 0.2187$\pm$0.0017 \\
			\bottomrule
		\end{tabular}
	}
\end{table*}

\textbf{(3) CR} is defined as the ratio of actual frame length to the byte count of the uncompressed reference point cloud
\begin{equation}
	\mathrm{CR}
	=
	\frac{
		B_{\mathrm{frame}}
	}{
		B_{\mathrm{ref}}
	}
	\times 100\%,
	\label{eq:cr}
\end{equation}
where \(B_{\mathrm{frame}}\) is the measured frame length including the frame header, and \(B_{\mathrm{ref}}\) is the storage cost of reference point cloud coordinates represented as 32-bit single-precision floating-point numbers
\begin{equation}
	B_{\mathrm{ref}}
	=
	M_{\mathrm{in}}\times 3\times 4.
\end{equation}
A lower CR indicates better performance.

\textbf{(4) Semantic mean Intersection-over-Union (mIoU).} The predicted category of the nearest reconstructed point is assigned to each reference point, and the IoU for valid class $c$ is computed as
\begin{equation}
	\mathrm{IoU}_c
	=
	\frac{
		\mathrm{TP}_c
	}{
		\mathrm{TP}_c+\mathrm{FP}_c+\mathrm{FN}_c
	},
	\label{eq:iou}
\end{equation}
The semantic mIoU is given by
\begin{equation}
	\mathrm{mIoU}
	=
	\frac{1}{C_{\mathrm{eval}}}
	\sum_{c=1}^{C_{\mathrm{eval}}}
	\mathrm{IoU}_c,
	\label{eq:miou}
\end{equation}
where \(C_{\mathrm{eval}}=13\), and unlabeled classes are excluded from computation. A higher mIoU indicates better performance. This metric only measures semantic preservation in reconstruction and does not represent downstream segmentation performance.

\textbf{(5) Statistical specification.} Test metrics are averaged over all test frames. For random channel experiments, three channel resampling runs are performed on the same model.

\subsection{Main Results}\label{subsec:main}

Table \ref{tab:main} compares the proposed method, zero-bit decoding and equal-byte coordinate direct transmission. The zero-bit baseline feeds the average codeword of the training set into the same decoder. For coordinate direct transmission, 898 coordinate points represented as 16-bit signed integers are retained via farthest point sampling and compressed via Zlib, yielding an average frame length of 5353 B with no semantic labels transmitted. All three schemes are evaluated on 500 test frames.

Compared with the zero-bit baseline, the proposed method reduces CD by 98.56\% and increases F@0.01 and mIoU by 83.7-fold and 67.2-fold respectively, indicating that codewords carry frame-specific information. Under comparable byte budgets, coordinate direct transmission achieves better geometric metrics, with a CD approximately 1/6.86 that of the proposed method. The proposed method reconstructs 9216 labeled points, reflecting a trade-off between geometric fidelity and semantic capacity. The proposed method achieves F-scores of \(0.6448\pm0.0008\) and \(0.9395\pm0.0006\) at thresholds of 0.02 and 0.05 respectively, indicating decent meter-level contour recovery but limited local details at approximately the 0.5 m scale.

\begin{figure}[t]
	\centering
	\includegraphics[width=\columnwidth,keepaspectratio]{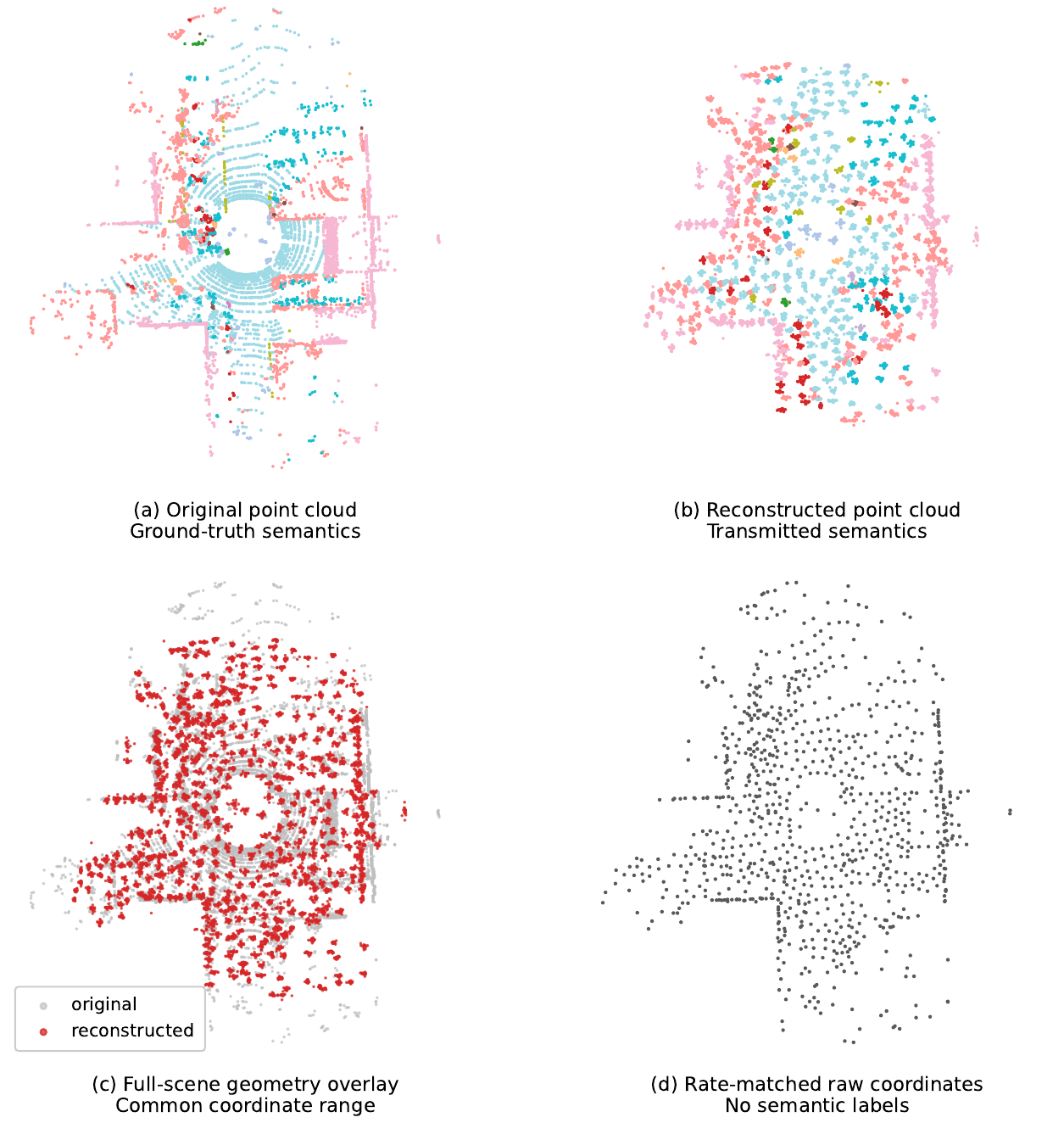}
	\caption{Qualitative results of a representative frame with CD closest to the per-frame median in the test set.
		(a) Reference point cloud colored by ground-truth semantic labels;
		(b) Reconstructed point cloud and its semantic labels from the proposed method;
		(c) Full-scene overlay of the reconstructed point cloud (red) and the reference point cloud (gray) within a unified coordinate range;
		(d) Result of equal-byte coordinate direct transmission.}
	\label{FigQual}
\end{figure}

Figure \ref{FigQual} shows that the proposed method only recovers the approximate spatial extent and partial category distribution of the scene, with noticeable differences in scanline structure and local point positions compared with the reference point cloud. Coordinate direct transmission yields more accurate geometric positions but carries no semantic labels.

\begin{table}[htbp]
	\centering
	\caption{Ablation of loss terms.}
	\label{tab:abl}
	\footnotesize
	\begin{tabular}{c c c c c}
		\toprule
		\(\beta\) & \(\gamma\) & CD & F@0.01  & mIoU\\
		\midrule
		0.7 & 0.2 & 0.001748 & 0.2903 & 0.2177 \\
		0   & 0.2 & 0.002065 & 0.2117 & 0.2051 \\
		0.7 & 0   & 0.001722 & 0.2942 & 0.0083 \\
		\bottomrule
	\end{tabular}
\end{table}
\subsection{Ablation Study}\label{subsec:ablation}
Table \ref{tab:abl} presents results obtained by separately removing the reconstruction loss \(L_{\mathrm{rec}}\) and semantic loss \(L_{\mathrm{sem}}\) under identical training and channel configurations, with their respective weights \(\beta\) and \(\gamma\).

Removing the reconstruction loss increases CD by 18.1\% and reduces F@0.01 by 27.1\%, indicating that \(L_{\mathrm{rec}}\) is indispensable for geometric recovery. Removing the semantic loss decreases mIoU from 0.2177 to 0.0083, a drop of approximately 96.2\%, demonstrating that \(L_{\mathrm{sem}}\) is critical for semantic information preservation. The full model strikes a balance between geometric quality and semantic retention. Checkpoints for the model with \(\gamma=0\) are selected based on validation CD, while those for remaining models are selected based on validation mIoU.

\subsection{Anchor Count and Transmission Overhead}\label{subsec:rate}
Table \ref{tab:rate} compares different anchor counts with fixed local shape code dimension and reconstruction point count. The corresponding expansion points per anchor are 36, 18 and 9 respectively.

\begin{table}[t]
	\centering
	\caption{Bit rate and reconstruction quality under different anchor counts.}
	\label{tab:rate}
	\footnotesize
	\begin{tabular}{c c c c c}
		\toprule
		Anchor count & Frame size (B) & CD  & F@0.01  & mIoU  \\
		\midrule
		256  & 2721  & 0.001916 & 0.2238 & 0.1812 \\
		512  & 5358  & 0.001748 & 0.2903 & 0.2177 \\
		1024 & 10242 & 0.001667 & 0.3821 & 0.2450 \\
		\bottomrule
	\end{tabular}
\end{table}

As the anchor count increases, both geometric and semantic metrics improve, while the bitrate overhead approximately doubles. Increasing the anchor count from 512 to 1024 raises the average frame length by 91.2\% but reduces CD by only 4.7\%. Considering both transmission overhead and reconstruction quality, 512 anchors are adopted as the default configuration in this work.

\begin{table}[t]
	\centering
	\caption{Channel sensitivity under the frame availability model. CD uses normalized coordinates.}
	\label{tab:channel}
	\footnotesize
	\resizebox{\columnwidth}{!}{%
		\begin{tabular}{c c c c @{\hspace{1.6em}} c c c c}
			\toprule
			SNR (dB) & CD & F@0.01 & mIoU & Additional packet loss (\%) & CD & F@0.01 & mIoU \\
			\midrule
			0  & 0.0643 & 0.0025 & 0.0005 & 0  & 0.0022 & 0.2691 & 0.2406 \\
			5  & 0.0643 & 0.0025 & 0.0005 & 5  & 0.0073 & 0.2455 & 0.2078 \\
			10 & 0.0361 & 0.1224 & 0.1145 & 10 & 0.0091 & 0.2369 & 0.2016 \\
			15 & 0.0036 & 0.2629 & 0.2215 & 20 & 0.0142 & 0.2127 & 0.1867 \\
			20 & 0.0022 & 0.2691 & 0.2406 & 50 & 0.0345 & 0.1290 & 0.1202 \\
			30 & 0.0022 & 0.2691 & 0.2406 &    &        &        &        \\
			\bottomrule
		\end{tabular}
	}
\end{table}
\subsection{Channel Sensitivity and Processing Latency}\label{subsec:channel}
Table \ref{tab:channel} examines channel sensitivity over 128 frames. The additional packet loss rate is set to 0 when sweeping SNR, and the SNR is fixed at 20 dB when sweeping packet loss rate.

Both low SNR and high packet loss rate significantly degrade frame availability. When SNR increases from 10 dB to 15 dB, CD drops from 0.0361 to 0.0036. When the additional packet loss rate rises from 0 to 50\%, CD increases from 0.0022 to 0.0345 and mIoU decreases from 0.2406 to 0.1202.

The model has a total of 2.41 M parameters, among which 2.18 M belong to the inference pipeline. On an RTX 4090 with a batch size of 1, the average latency over 20 measurements is 72.70 ms per frame, including 64.67 ms for preprocessing. The latency without preprocessing is 8.03 ms per frame, and the core pipeline of encoding, channel simulation and decoding takes 7.73 ms per frame.

\subsection{Multi-View Fusion Validation}
Six time windows are selected from the test sequence, each containing 3 frames with a 2 s interval, and aligned according to poses. Both single-frame reconstructions and fused results are evaluated against the union of the three pose-aligned preprocessed point clouds.

\begin{table}[htbp]
	\centering
	\caption{Fusion results over 6 time windows (mean \(\pm\) standard deviation across time windows; all schemes use the same aligned union reference).}
	\label{tab:fusion}
	\footnotesize
	\resizebox{\columnwidth}{!}{%
		\begin{tabular}{l c c c c}
			\toprule
			Scheme & CD (m\(^2\)) & F@0.1 m & F@0.25 m & F@0.5 m \\
			\midrule
			Best single view per window & 5.526$\pm$1.208 & 0.0140$\pm$0.0029 & 0.1067$\pm$0.0204 & 0.3180$\pm$0.0504 \\
			\textbf{\(\Psi_{\text{fuse}}\) fusion} & \textbf{2.363$\pm$0.344} & \textbf{0.0244$\pm$0.0042} & \textbf{0.1708$\pm$0.0254} & \textbf{0.4586$\pm$0.0447} \\
			\bottomrule
		\end{tabular}
	}
\end{table}

Fusion reduces the average CD from 5.526 to 2.363 m\(^2\) and increases F@0.5 m from 0.3180 to 0.4586, corresponding to improvements of 57.3\% and 44.2\% respectively. On average, 12.1\% of candidate point pairs satisfy the consistency criterion. As SemanticPOSS is a single-vehicle continuous sequence, this result only validates asynchronous multi-view fusion and does not represent real simultaneous multi-vehicle sensing.

\section{Conclusion}\label{Sec5}
A spatial-semantic anchor communication system for IoV point clouds was proposed in this paper, which realized joint transmission of local geometry and semantic categories under explicit coordinate constraints.
Experiments on SemanticPOSS demonstrated that digital frames with an average size of 5358 B accounted for only 4.84\% of the uncompressed coordinate overhead, and reduced the normalized CD by 98.56\% relative to the zero-bit baseline. This confirmed that anchor codewords effectively carry frame-specific information.
Loss ablation validated the complementary effects of geometric constraints and semantic supervision. Equal-byte coordinate direct transmission still yielded lower geometric errors, reflecting a trade-off between semantic capacity and geometric fidelity.
Asynchronous multi-view fusion further reduced the metric CD by 57.3\%. The local simulation processing latency on a single RTX 4090 was 72.70 ms per frame.

\section*{CRediT authorship contribution statement}
\textbf{Ruixing Ren:} Investigation, Methodology, Software, Writing-original draft, Visualization
; \textbf{Junhui Zhao:} Writing-review \& editing, Supervision, Validation, Project administration, Funding acquisition; \textbf{Wei Zhao:} Writing-review \& editing, Validation; \textbf{Gaoxin Zhang:} Writing-review \& editing, Validation; \textbf{Xiaoming Wang:} Writing-review \& editing, Validation.

\section*{Acknowledgements}
This work was supported in part by National Natural Science Foundation of China under  Grant U25B2007, in part by the Fundamental Research Funds for the central Universities under Grant 2025JBZX060, in part by National Engineering Research Center of System Technology for High-Speed Railway and Urban Rail Transit under Grant 2024YJ255.

\section*{Declaration of competing interest}
The authors declare that they have no known competing financial interests or personal relationships that could have appeared to influence the work reported in this paper.

\section*{Data availability}
The experiments in this work are conducted on the SemanticPOSS dataset, which is publicly available for non-commercial research purposes at https://poss.pku.edu.cn/sem\\anticposs.html.

\bibliographystyle{elsarticle-num} %为文献的格式类型
%\balance
\bibliography{MyRefs}

%~~~\\
%~~~\\

% References are generated from MyRefs.bib using elsarticle-num.bst.

\end{document}